\RequirePackage{fix-cm}
\documentclass[smallextended]{svjour3}       
\smartqed  
\usepackage{graphicx,natbib,amsmath,xcolor,hyperref}
\newcommand{\diff}{\text{\rm d}}

\newcommand{\Bxi}    {\ensuremath{\boldsymbol\xi}}
\newcommand{\Beta}   {\ensuremath{\boldsymbol\eta}}

\newcommand{\Bb}{{\boldsymbol{\mathnormal b}}}

\newcommand{\Be}{{\boldsymbol{\mathnormal e}}}
\newcommand{\Bf}{{\boldsymbol{\mathnormal f}}}

\newcommand{\Bu}{{\boldsymbol{\mathnormal u}}}

\newcommand{\Bx}{{\boldsymbol{\mathnormal x}}}
\newcommand{\By}{{\boldsymbol{\mathnormal y}}}

\def\makeheadbox{\relax}

\newcommand \MZ [1] {\bgroup\noindent[\textcolor{blue}{\textbf{MZ}: #1}]\egroup\ignorespacesafterend}
\newcommand \SG [1] {\bgroup\noindent[\textcolor{violet}{\textbf{SG}: #1}]\egroup\ignorespacesafterend}

\begin{document}

\title{Peridynamics based model of anticrack-type fracture in brittle foams
}


\author{Shucheta Shegufta         \and
        Michael Zaiser 
}


\institute{S. Shegufta and M. Zaiser\at
           Friedrich-Alexander Universit\"at Erlangen-N\"urnberg \\
           Department of Materials Science, WW8-Materials Simulation\\
           Dr.-Mack Strasse 77, 90762 F\"urth, Germany
           \email{shucheta.shegufta@fau.de}           
}

\date{Received: date / Accepted: date}

\maketitle

\begin{abstract}
A particular failure mode of highly porous brittle materials consists in the propagation of cracks under uniaxial compressive loads. Such 'anticracks' have been observed in a range of materials, from snow and porous sandstone to brittle foams. Here we present a computational model for the formation and propagation of anticrack-type failure in porous materials within the general computational framework of bond-based peridynamics. Random porosity is represented, on a scale well above the characteristic pore size, by random bond deletion (dilution disorder). We apply our framework to experimental data on anticrack propagation in silicate foams.  
\keywords{fracture \and peridynamics \and anticracks}
\end{abstract}

\section{Introduction}
\label{intro}
In the regime of brittle or quasi-brittle behavior, highly porous materials allow for an at first glance counter-intuitive failure mode:  failure may occur under loading conditions when the principal stresses are purely negative, a situation precluded by conventional wisdom. In case of uniaxial mode-I loading, such failure implies on the macroscopic scale a negative crack face displacement ('anticrack'), which is possible because of the formation of a damage zone around the crack faces in which the material porosity strongly decreases while microscale contact of the opposing crack faces is not yet established. Such behavior has been observed in a range of porous materials, for example sandstone \citep{mollema1996compaction,sternlof2005anticrack}, snow \citep{heierli2008anticrack,heierli2008failure,gaume2018dynamic,adam2024fracture}, and glass foams \citep{heierli2012anticrack}. Simulation studies have reported similar behavior in cemented granular materials \citep{yamaguchi2020failure}. A similar phenomenon in geosciences are pressure solution surfaces \citep{fletcher1981anticrack} where negative crack face displacement is made possible by pressure induced dissolution and removal of material. We note that a second use of the term 'anticrack', namely to denote rigid lamellar inclusions in a soft matrix \citep{dundurs1989green}, is not directly pertinent to the present work. 

Computational and theoretical models of anticrack-type failure have used a range of methods, including analytical fracture mechanics arguments \citep{sternlof2005anticrack,heierli2008anticrack}, discrete element method (DEM) \citep{ritter2020microstructural,yamaguchi2020failure}, and continuum modelling. Of particular interest for the present work are elastoplastic models that allow for volume reduction as developed for snow by \citet{barraclough2017propagating},\citet{gaume2018dynamic},\citet{lowe2020snow} and implemented for dynamic anticrack propagation in snow using the material point method by \citet{gaume2018dynamic}. This implementation is of particular interest since it naturally takes into account dynamic fragmentation phenomena subsequent to the release of an avalanche by anticrack propagation. More generally speaking, the handling of complex fracture geometries and the contact between fracture surfaces is an indispensable feature of anticrack modelling (where ultimate contact of fracture surfaces is intrinsically part of the problem), and particle based methods like DEM or MPM are natural candidates for such problems. 

Here we use another mesh-free, particle-based simulation approach that allows for dealing with spontaneous emergence and self-organized development of fracture surfaces, namely bond-based peridynamics enhanced by contact forces. We parameterize and validate this approach by reference to the excellent experimental study of \citet{heierli2012anticrack} who investigated anticrack-type failure in brittle foam glass. A description of the model is provided in Section \ref{sec:peridyn} alongside a parameterization of the model for the foam glass material of \citet{heierli2012anticrack}. Results for different loading modes are presented in Section \ref{sec:results} and compared with experimental findings, along with post-failure behaviour of the material. The results are summarized in Section \ref{sec:conclusions}

\section{Model formulation and parameterization}
\label{sec:peridyn}

\subsection{Bond based peridynamics: basic model}
\label{subsec_Peri_model}

We use a bond-based peridynamic model as originally defined by \citet{silling2010peridynamic}. Since the experiments of \citet{heierli2012anticrack} can be envisaged as plane strain deformation, we consider quasi-two-dimensional systems where the displacement field is a two-dimensional vector $\Bu(\Bx) = \By(\Bx) - \Bx$ where $\By(\Bx)$ is the current location of the point with material coordinates $\Bx$.  The force balance equation for the point $\Bx$ is given by
\begin{equation}
	\rho(\Bx) \ddot{\Bu}(\Bx) = \int_{{\cal H}_{\Bx}} \Bf \left(\Bx,\Bx^{\prime} \right) \diff \Bx^{\prime} + \Bb(\Bx) ,
    \label{eq:forcebalance}
\end{equation}
where $\Bf \left( \Bx,\Bx^{\prime} \right)$ is the pair force between $\Bx^{\prime}$ and $\Bx$, $\Bb$ is a body force field, and interactions are restricted to a family ${\cal H}_{\Bx}$ which we take to be a circle of radius $\delta$, the so-called horizon, around $\Bx$, $\left\vert \Bx - \Bx^* \right\vert \le \delta \; \forall \; \Bx^* \in {\cal H}_{\Bx}$. 

The pair force is specified constitutively. We introduce the notations $\Bxi \left(\Bx,\Bx^{\prime} \right) = \Bx^{\prime} - \Bx$ and $\Beta \left(\Bx,\Bx^{\prime} \right) = \Bu \left(\Bx^{\prime} \right) - \Bu \left(\Bx \right)$. The pair force is then taken linearly proportional to the bond stretch $s$, and pointing in the direction of the vector $\Be_u$ connecting both points in the current configuration:
\begin{equation}
	\Bf \left(\Bx,\Bx^{\prime} \right)= c \left(\Bxi \right) s \Be_u \left(\Bxi,\Beta \right) \quad,\quad s = \frac{\left\vert \Beta + \Bxi \right\vert - \left\vert \Bxi \right\vert}{|\Bxi|}\quad,\quad \Be_u \left(\Bxi,\Beta \right) = \frac{\Beta + \Bxi}{ \left\vert \Beta+\Bxi \right\vert }
\end{equation}
Here $c(\Bxi)$ is the so-called bond micro-modulus which for an isotropic bulk material depends on the bond length $\xi = |\Bxi|$ only. In the following we use for simplicity a constant micro-modulus, $c(\xi)=c_0$, which we choose such that the behavior of the material under homogeneous small deformations matches an isotropic linear-elastic material of bulk modulus $k$, hence $c_0 = 12k/\pi \delta^3$. The bond energy can then be written in terms of the bond length $\xi$ and bond stretch $s$ as
\begin{equation}
	e \left(\Bx,\Bx^{\prime}\right) = 	e\left(\Bx^{\prime},\Bx\right) = \frac{c_0}{2} s^2 \xi.
\end{equation}

Elastic-brittle behavior is normally introduced by defining a critical bond stretch $s_{\rm c}$ in such a manner that, once $s > s_{\rm c}$, the bond fails: in this case, its micro-modulus is irreversibly set to zero. The elastic energy $\frac{c_0}{2} s_{\rm c}^2 \xi$ of the failed bond is then converted into defect energy $E_{\rm d}$; for our later considerations it is irrelevant whether the defect energy is considered as some kind of microscopically stored, non recoverable internal energy, or simply as heat. In the present work we consider the mechanical behavior of a microscopically brittle material, namely closed-cell glass foams, for which a modified failure rule is needed as two failure thresholds are required: Bond failure can occur either for $s < -s_{\rm cb}$, representing compressive buckling of the cell walls with immediate brittle failure, or for $s > s_{\rm tf}$, representing tensile failure. 

For numerical implementation, we consider discretization of the continuous medium in terms of a regular grid of collocation points with spacing $\Delta$. This numerical implementation converges to the limit of Eq. (\ref{eq:forcebalance}) as $m = \delta/\Delta \to \infty$
(so-called $m$ convergence). 

\subsection{Porosity and elastic properties}

To mimic randomly distributed voids, we build upon the model of \citet{chen2019peridynamic}. These authors proposed a mesoscale 'intermediately-homogenized' (IH) peridynamic model and applied it to wave propagation in glass foams. Their basic idea is to formulate a model which does not resolve single voids or other microstructural details, but nevertheless accounts for the porosity dependency of elastic properties as well as for microstructural disorder typically associated with large degrees of porosity. For a material with porosity $\phi$ and critical porosity $\phi_{\rm c}$ (the porosity where the material loses cohesion), this is done by the simple expedient of randomly deleting a fraction  $d_{\phi} = 1 - (1 - \phi/\phi_{\rm c})^2$ of all bonds. For the limit of sufficiently large $m$, \citet{chen2019peridynamic} demonstrated that elastic moduli $E_{\phi}$ and wave propagation speeds in the IH model converge to those of a fully homogenized (FH) model where all bonds are retained but the bond micro-modulus $c_0$ (thus, equivalently, the macroscopic modulus $E_0$) is reduced by the constant factor $(1 - \phi/\phi_{\rm c})^2$. They also show that their model can well account for the porosity dependence of elastic modulus $E_{\phi}$ and wave propagation velocity in foam glass if one sets $\phi_{\rm c} = 1$, corresponding to the proportionality $E_{\phi} \propto E(\rho_{\phi}/\rho_0)^2$ as $\rho_\phi = \rho_0(1-\phi)$. 

Because of the random nature of the bond deletion process, bond deletion not only reduces the effective elastic modulus but also introduces fluctuations in local stiffness which, as we shall demonstrate, increase with increasing porosity. \citet{chen2019peridynamic} claim that, as a consequence of this randomness, the intermediately-homogenized model can account for the effects of strength fluctuations in highly porous materials. To analyze this claim and understand the nature of fluctuations in the model, some elementary statistical considerations are helpful. 

The maximum number of bonds connecting to a collocation point can for a 2D model (and disregarding subtleties in treating pairs whose distance is close to $\delta$) in the limit of sufficiently large $m$ be estimated as $N_{0} \approx \pi m^2$. Each of these bonds is retained with independent probability $p_{\phi} = (1 - \phi/\phi_{\rm c})^2$. Thus, the number of bonds connecting to any collocation point is a binomial distributed random variable. Its mean and standard deviation are given by
\begin{equation}
    \mu_N = N p_{\phi} \quad,\quad
    \sigma_N = \sqrt{N p_{\phi}(1-p_{\phi})}
    .
\end{equation}
As a measure of disorder inherent in the distribution (relative variations in local modulus), we may use the coefficient of variation 
\begin{equation}
    C_{\rm V} = \frac{\sigma_N}{\mu_N} 
     = \sqrt{\frac{1}{N_0} \frac{d_{\phi}}{p_{\phi}}}
     \approx{\frac{1}{m\left(1-\frac{\phi}{\phi_{\rm c}}\right)}}
    \sqrt{\frac{1-\left(1-\frac{\phi}{\phi_{\rm c}}\right)^2}{\pi}}.
\end{equation}
This expression shows that fluctuations diverge as the porosity approaches its critical value, an idea that is consistent with the critical nature of a percolation transition. However, we also see that the coefficient of variation goes to zero in the limit of large $m$, which implies that in this limit the IH model converges to a fully homogeneous model: The ability of the model to represent microstructural disorder is contingent on the choice of the discretization parameter $m$. This undesirable feature has not been recognized in the study of \citet{chen2019peridynamic}. Moreover, we note that in general disorder is not only a function of porosity: Compare a closed-cell microstructure with perfectly uniform cell sizes akin to a honeycomb pattern with a random Apollonian packing \citep{dodds2002packing} with an extreme scatter of cell sizes. 

To allow for independent variation of the mean elastic modulus and the scatter of local strength values, and to avoid the undesirable $m$ dependency of the latter, we modify the model of \citet{chen2019peridynamic}. We consider situations where the macroscopic elastic modulus depends on relative density as $E_{\phi} = E_0(1-\phi)^a$ but we achieve this result by using a randomization approach where we rescale the bond micro-modulus by a factor $\eta_{\phi,CV}$ and at the same time randomly delete a fraction $1-p$ of bonds. As a result, the effective elastic modulus changes by 
\begin{equation}
E = E_0 \eta_{\phi,CV}p_{\phi,CV} := E_0(1-\phi)^a
\end{equation}
and at the same time the coefficient of variation is
\begin{equation}
CV = \frac{1}{m} \sqrt{\frac{1-p}{\pi p}}
\end{equation}
From these two equations the factors $\eta$ and $p$ can be determined:
\begin{equation}
p_{\phi,CV} = \frac{1}{1 + \pi^2 m^2 CV^2} \quad,\quad
\eta_{\phi,CV} = (1-\phi)^a (1 + \pi m^2 CV^2)
\end{equation}
This has the interesting consequence that, for typical values of $m$ and $CV$, the factor $\eta$ may be bigger than 1, which implies an anti-homogenization approach. For example, set $\phi = 0.4$, $m=4$, $a=2$ and $CV=0.25$ then we find $\eta=1.49,p=0.24$, thus, the desired disorder is achieved by deleting an excessive fraction of bonds while strengthening the rest (anti-homogenization). 

\subsection{Contact law}
Upon compression of highly porous materials, local failure may lead to sudden compaction accompanied by a loss of strength. This is counter-acted by contact forces between the debris which need to be included to correctly describe the behavior of porous materials under compressive loads. For this purpose, the so-called short-range force approach is adapted where a repulsive force $\Bf_s$ acts between material points if they get within a certain distance $r_s$ to each other. We follow, with a slight modification, the approach of \citet{kamensky2018contact}. The force balance equation with contact forces now reads in absence of body forces
\begin{equation}
	\rho(\Bx) \ddot{\Bu}(\Bx) = \int_{{\cal H}_{\Bx}} \Bf \left(\Bx,\Bx^{\prime} \right) \diff \Bx^{\prime} + \int_{\cal V \backslash \cal V_{\rm self}(\Bx)} \Bf_{\rm c} \left(\Bx,\Bx^{\prime} \right) \diff \Bx^{\prime}
    \label{eq:forcebalancecontact}
\end{equation}
Contact of the point with material coordinate $\Bx$ is allowed with any other material point $\Bx'$ within the system volume ${\cal V}$, excluding a 'self interaction' domain $V_{\rm self}$ which we take to be a circle of radius $r_{\rm self}$. The contact force density is given by
\begin{equation}
    \Bf_{\rm c} = 
    \begin{cases}
    \displaystyle
    c_{\rm c} \frac{|\Beta + \Bxi| - r_{\rm c}}{r_{\rm c}} \Be_{\Bu}(\Bxi,\Beta) & \text{if }\left\vert \Beta + \Bxi \right\vert \le r_{\rm c} \\
    0 & \text{otherwise.}
    \end{cases}
\end{equation}
This expression describes a repulsive radial force between contacting points that have, in the current configuration, separation less than $r_{\rm c}$. It diverges as the distance between the points goes to zero, preventing interpenetration of material. For contact distances close to $r_{\rm c}$, linear-elastic behavior is recovered where the contact stiffness $c_{\rm c}$ must be adjusted to match the compressive modulus of the contacting debris. 

The radius of exclusion of self-interaction $r_{\rm self}$ must be less than the radius of contact $r_{\rm c}$. In the present physical context, both radii must be adjusted to match the stress-free 'collapse strain' that accrues after bonds break under compressive load and before the resulting debris get into contact.

\subsection{Parameterization of the model}

We consider a 2D system with plane stress loading, which implies because of the central forces that Poisson's number $\nu = \frac{1}{3}$ . For discretization we use a square planar grid of collocation points with a spacing $d = 1$ mm, which also formally defines the 'thickness' of the system in the third dimension. The radius of the horizon is chosen as $\delta = 4.001$ mm. We have verified that reducing this value to $2.001$ mm has no significant influence on the simulation results. 

Material properties of the matrix are chosen to reflect those of soda-lime silicate glass as given by \citet{kilinc2015mechanical}, with a Young's modulus $E_{\rm m} = 74$ GPa, and density $\rho = 2620$ kg/m$^3$. To reflect the $95\%$ porosity of the foam, we set $\eta=0.0075, p=0.33$, by choosing a $CV=0.2$ and setting the density scaling exponent $a=2.43$, leading to an elastic modulus of the foam material of $E_{\rm f} = 48.12$ MPa. The specific fracture energy $G_{\rm c}$ for compressive Mode-I failure is taken to be $125$ Jm$^{-2}$, which leads to a critical bond stretch for compression $s_{\rm c} = 0.0126$. This value allows us to recover the experimentally observed area under the stress-strain curve of a specimen without initial flaws with a compressive failure strain of 0.67\% (see Figure  \ref{fig:damagePattern}, top). 

The tensile Mode I specific fracture energy is chosen to be be $12.5$ Jm$^{-2}$, which, in accordance with \citet{heierli2012anticrack}, is an order of magnitude smaller than its compressive counterpart and corresponds to a critical stretch $s_{\rm t} = 0.004$. 

To calibrate the contact parameters, we assume that under confined conditions, the porous material and debris will have the same elastic properties. Here, we define debris as a material point that has no peridynamic bonds with other material points, in other words, a material point whose local damage is one and which only interacts with other material points by contact forces. For $r_{\rm c} = 0.8$, we recover the elastic properties of the intact porous material by setting $c_{\rm c} = 7.14$N/mm$^{-6}$.

\section{Simulation results and comparison with experiment}
\label{sec:results}
We now apply the parameterized model to simulate the experiments reported by \citet{heierli2012anticrack}, who studied the fracture behavior of cellular foam glass under compressive loading. Experiments were conducted on samples of $200\times60\times20$ mm$^3$ size, cut from larger plates, and loading was applied using two thick aluminum plates placed parallel to the top and bottom surfaces. Since cutting of the samples weakened the cells near the surface, top and bottom surfaces were reinforced by coating them with epoxy resin. Samples without coating were also tested for comparison. 

\begin{figure}[tbh]
  \includegraphics[width=\textwidth]{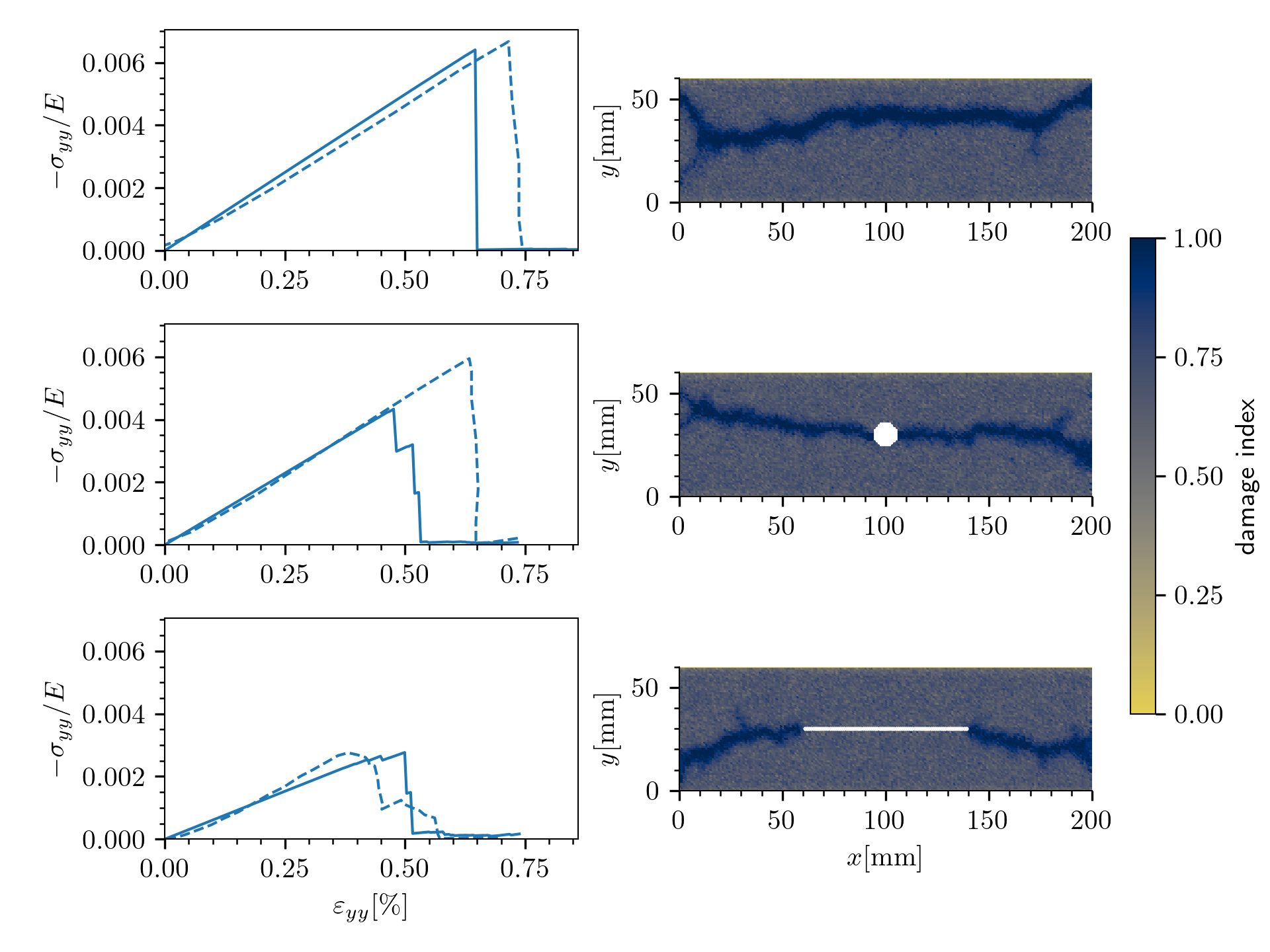}
\caption{Stress-strain curves and damage patterns of three different cases obtained using the homogenized PD model; top: rectangular plate without pre-crack, middle: plate with a central hole; and bottom: plate with a central notch}
\label{fig:damagePattern}       
\end{figure}

In the first of the three reported experiments, the top layer is compressed at a rate of $\dot{u} = 2$ mm min$^{-1}$. We apply the same loading rate to a 2D model of $200\times60$ mm$^2$ with additional, clamped boundary layers of thickness $\delta$ at the top and bottom to mimic the effect of epoxy coating. Results (stress-strain curves and damage patterns at $0.72$\% strain) are shown in figure \ref{fig:damagePattern}. The elastic moduli of the experimental specimens show a very significant scatter. Therefore, in order to facilitate comparison between simulation and experiment, we normalize all stresses by the elastic modulus, defined as the maximum slope of the stress-strain curve during the initial loading stage. 

As in the experiments, the model responds in an approximately linear-elastic manner until a compressive strain of about 0.7\%, at which point the load drops suddenly to a level near zero. This load drop is accompanied by pronounced localization of both compressive strain and bond damage in a zone spanning the system approximately perpendicular to the load axis ('compaction band'). While the orientation of the compaction band perpendicular to the load axis is similar to the experimental observations, the same is not true for the 'vertical' location of the band: In the experiments, damage always initiates at the interface between the epoxy surface coating and the bulk material, whereas in the present simulations, the location of the compaction band varies from specimen to specimen. 

The second experiment follows the same loading scheme, but this time the specimen contains a central hole of $14$ mm diameter. In this case, fracture nucleates at the sides of the hole, corresponding to the locations where classical continuum mechanics predicts the maximum compressive stress. According to a well-known result by Kirsch for a circular hole in an infinite plate under uni-axial remote loading, the stress at these locations is expected to be three times the axial stress. These stress concentrations lead to a reduction in the peak stress and thus of the critical strain required to trigger fracture propagation. Starting from the sides of the initial hole, compaction bands then propagate perpendicular to the loading direction, causing the load to drop to near zero. The behavior observed in the experiment matches this scenario, even though the reduction in peak stress is less pronounced than in the simulation. 

The third type of experiment involves a specimen with a central notch of $80$ mm length and $2$mm thickness. Loading conditions are similar as in the previous cases. We introduce the notch by using a 'bond filter', which deletes all bonds crossing the specified plane in the model. The model is able to capture the fracture behavior observed in the experiment, where crack propagation is preceded by  multiple small fracture events, indicated by a small load drop and a slight reduction of the slope of the stress strain curve. Then a deformation band propagates on one side of the notch to the surface, reducing the load to half. Shortly after, a second deformation band propagates rapidly from the other side of the notch and the load drops to zero. The critical loads and strain levels for fracture propagation in simulation and experiment are in good agreement (note that because of the presence of the central notch, the effective specimen cross section and thus the apparent elastic slope are reduced relative to the bulk sample). 

We may ask what parameters control the failure stress in our simulations. First, from the experiment with the circular hole we conclude that failure is not controlled by the maximum stress in the sample: The maximum stress at the left and right edges of the hole is expected to exceed the applied compressive stress by a factor of 3, however, this does not lead to a corresponding reduction of the sample failure stress, which decreases only by about 30\%. 

\begin{figure}[tbh]
\centering
  \includegraphics[width=.6\textwidth]{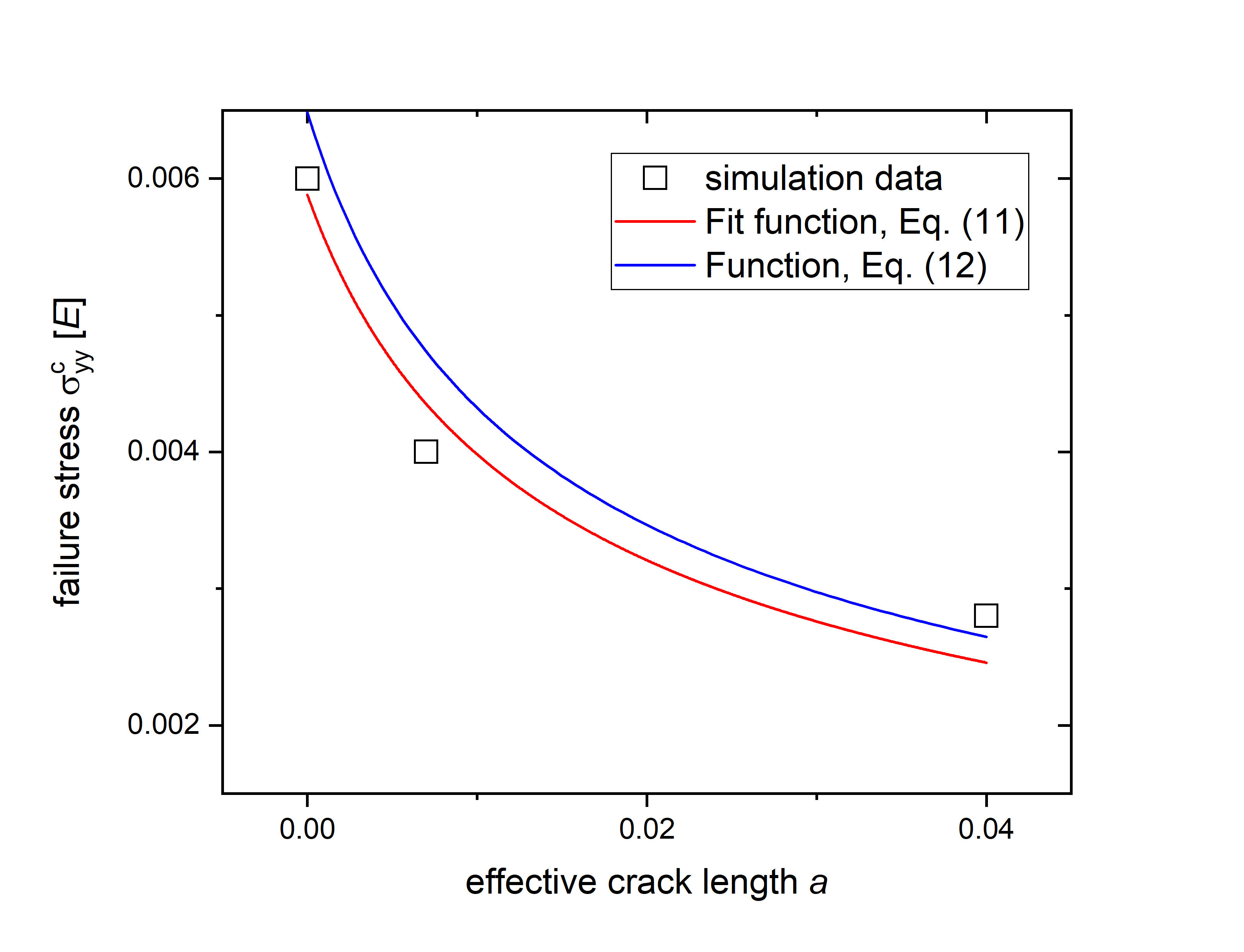}
\caption{Dependency of failure stress on flaw size; squares: simulation data, red line: best-fit curve according to Eq. (11), blue line: curve according to Eq. (12).}
\label{fig:sizeEffect}       
\end{figure}

A second conjecture is that the local stress concentration at the sides of the hole, or at the end of the notch, lead to formation of a crack-like flaw whose propagation is controlled by a Griffith-like energy criterion. Indeed, the dependency of failure stresses on flaw size can be well fitted by the phenomenological size effect law of \citet{bazant1984size},
\begin{equation}
\sigma_{yy}^{\rm c}= \frac{\sigma_{yy}^{\rm c}(0)}{\sqrt{1+\frac{a}{a_0}}}
\end{equation}
Setting the flaw size $a=0$ for the defect-free specimen, $a=7$mm (hole radius) for the specimen with hole, and $a=40$ mm (notch half length) for the specimen with notch, results in the data shown in figure \ref{fig:sizeEffect}, bottom,
where a best fit gives $\sigma_{yy}^{\rm c}(0) = 0.00588 E$ and $a_0 = 8.47$ mm (red curve in figure \ref{fig:sizeEffect}, bottom). Alternatively, we can represent the data in a parameter-free manner using the Griffith-like expression
\begin{equation}
\sigma_{yy}^{\rm c}= \frac{\sqrt{G E}}{\sqrt{\pi(a + 2 \delta)}}
\end{equation}
where we assume quasi-brittle failure with the fracture energy $G$ taken from our model parameterization and estimate the process zone size $a_0 \approx 2 \delta$ by the horizon diameter (blue curve in figure \ref{fig:sizeEffect}, bottom). The good agreement between theoretical curves and failure stress data supports an energetic scenario where the failure stress is controlled by the requirement that the stress must be high enough such that the energy release upon crack propagation matches the fracture energy. 

\subsection{Post-failure behavior: Cleavage}
To analyze the nature of the failure process that occurs under compressive loading once the critical stress is exceeded, we use two different loading protocols. First, we subject a notched specimen similar to case 3 first to compressive displacement and then, after the load drop, we unload the specimen and reverse load in tension. Figure \ref{fig:unloading} shows the loading scheme and snapshots of the damaged specimen and movie 1 in the supplementary video shows the entire damage evolution. We first compress the top layer up to $0.45$ mm, after which we reverse the displacement and displace the top layer upwards up to $1.8$ mm. In the compression regime, the material behaves similar to the case discussed above, where fracture initiates and propagates first at one side of the notch, dropping the load about halfway, and then from the other side, causing the load to drop to almost zero. During unloading to zero displacement the load goes to zero and during subsequent reverse loading, the material displays no resistance and the load remains at zero. Figure \ref{fig:unloading}d shows that, during reverse loading, the specimen separates along the main fracture path, indicating the material is completely broken across this plane. In other words, the failure that has led to the compressive load drop has cleaved the specimen: We are dealing with a compressive mode-I fracture. 

\begin{figure}[tbh]
  \includegraphics[width=\textwidth]{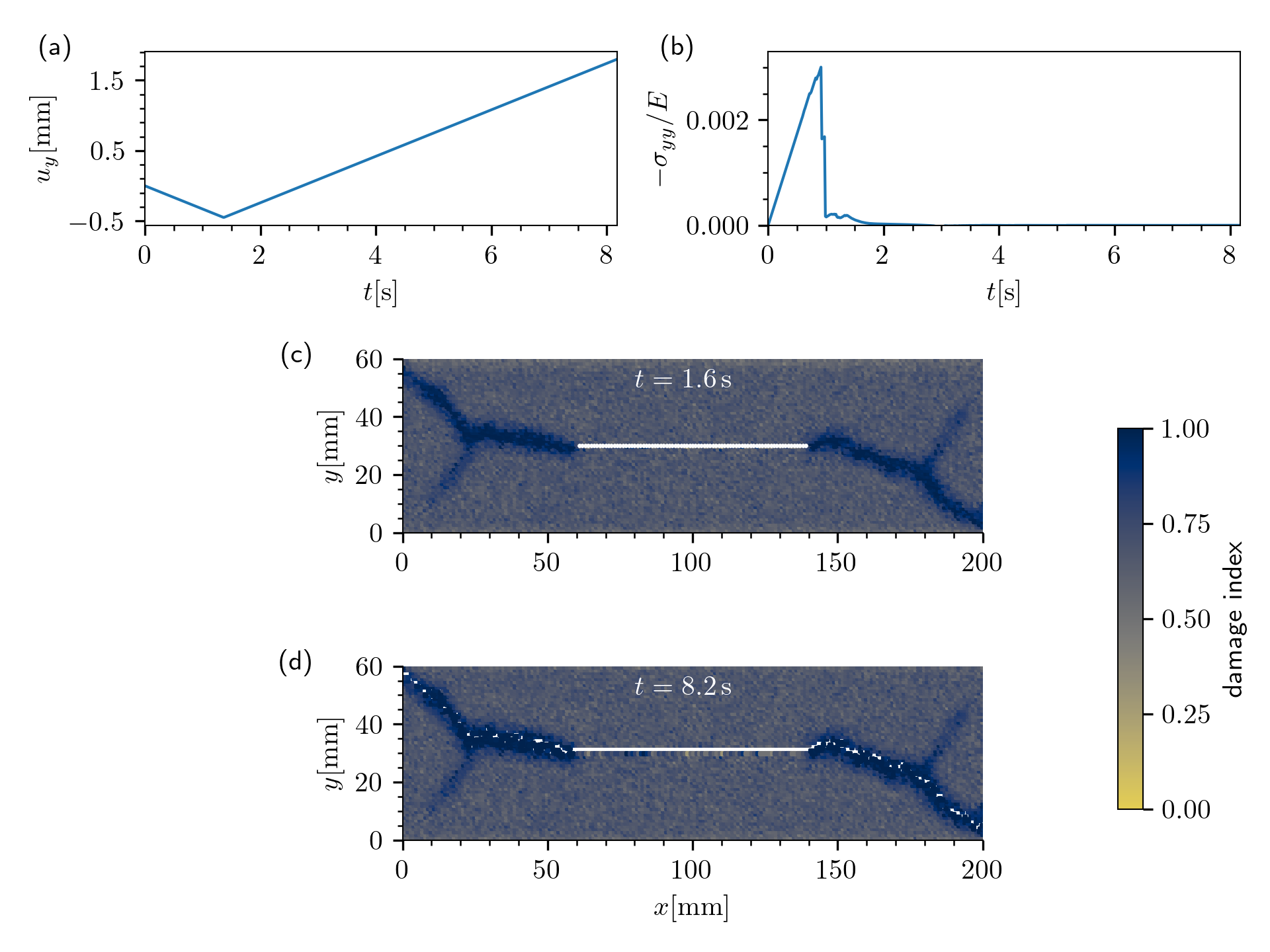}
\caption{Loading scheme illustrating compressive cleavage of the notched specimen; (a) applied displacement at the upper layer vs time, (b) Stress normalized with elastic modulus vs time, (c) damage pattern of the specimen at the onset of fracture and (d) after reverse loading}
\label{fig:unloading}       
\end{figure}

\subsection{Post-failure behavior: compaction}
\begin{figure}[tbh]
  \includegraphics[width=\textwidth]{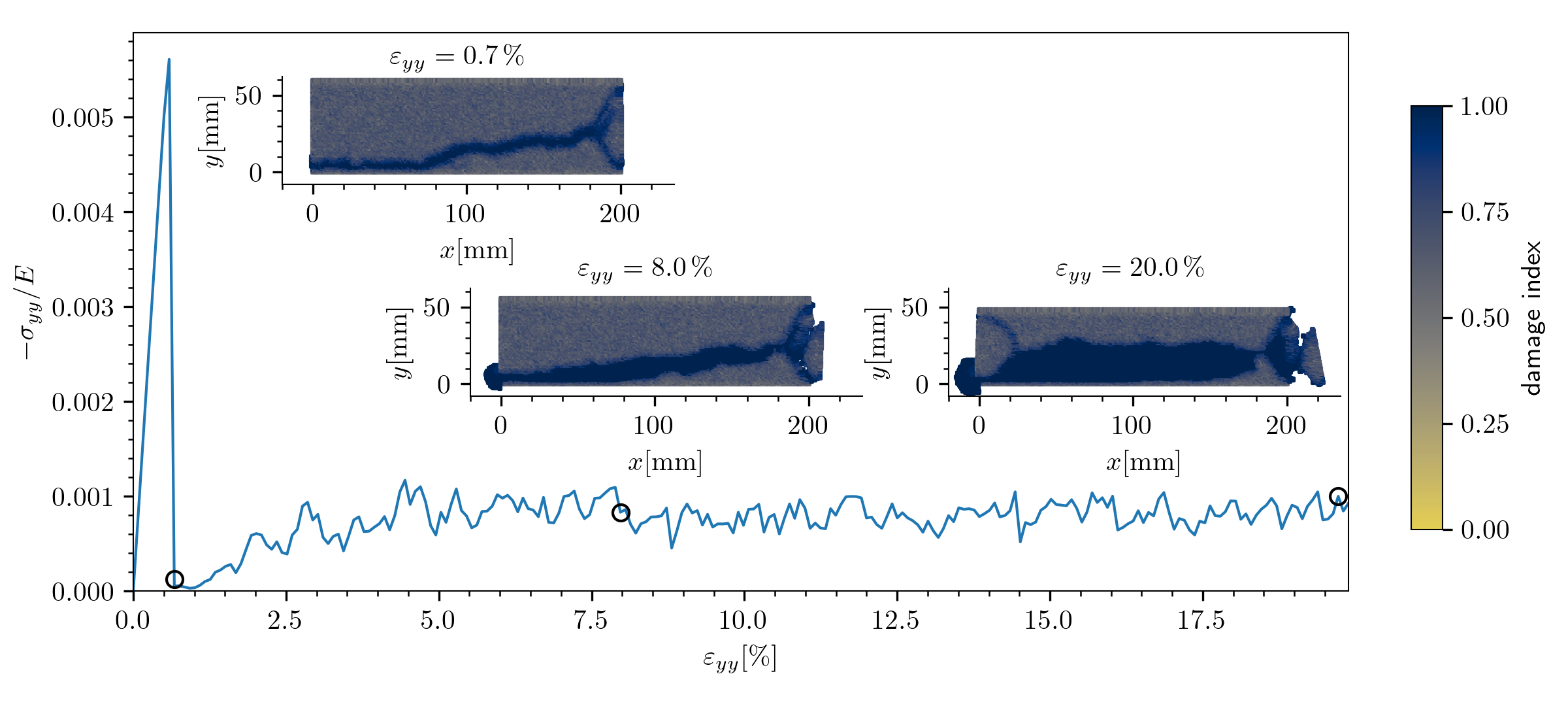}
\caption{Stress-strain curve and damage patterns for foam glass under compression; top row: stress normalized with elastic modulus vs strain; bottom row, from left to right: damage pattern of the specimen at 0.7\%, 8\% and 20\% strain.}
\label{fig:compaction}       
\end{figure}

We now focus on the compaction behaviour of the material by continuing the compressive loading to a compressive displacement of $12$mm. We consider an initially homogeneous specimen. Similar to the previous section, a compaction band initiates at a compressive strain of around $0.7$\% and propagates perpendicular to the loading direction across the specimen width, accompanied by a sudden drop of the load to near zero (first inset in figure \ref{fig:compaction}). Along the compaction band, bonds are almost completely broken and cohesion is lost. However, the compressive strain in the band cannot increase indefinitely since the resulting debris get into mutual contact and these contact forces prevent further compression within the band. Therefore we observe after the initial load drop an increase in the applied load which then stabilizes around a plateau level. Further compaction is not achieved by increasing the compaction strain within the band, but by widening of the band, which expands laterally in the direction of the loading axis ('vertical' direction in the figures showing damage patterns, see second and third insets in figure \ref{fig:compaction}, corresponding to 8\% and 20\% compressive strain). We may thus conclude that the plateau stress, which amounts to about one sixth of the peak load, corresponds to the stress required for the lateral expansion of the compaction band into undamaged material.
The formation and lateral expansion of compaction bands is a generic feature of porous materials. The mode of compression by lateral expansion of a compaction band which we observe here was for example reported and theoretically modelled for the case of snow, see work of Barraclough et. al. \cite{barraclough2017propagating}. Here we directly demonstrate the close relation between the formation of propagating compaction bands and loss of cohesion by compressive mode-I failure (anticrack formation).

\section{Discussion and conclusions}
In this work, we have introduced a homogenized model for the failure of porous media with internal disorder. We started from the IH model proposed by \citet{chen2019peridynamic}, who introduce disorder into a bond-based peridynamic model by random bond deletion. This approach has the drawback that the disorder becomes dependent on the discretization parameter $m$ and goes to zero in the limit of large $m$. In other words, the model does not achieve $m$ convergence and becomes more homogenized with increasing values of $m$.  We have addressed this problem by simultaneous bond deletion and re-scaling  of the micro-modulus of the remaining bonds, which is a function of both porosity and $m$. The advantage of the modified method is twofold; first, we are now able to maintain a desired disorder in the system irrespective of $m$, and second, the new method allows for more flexibility in choosing the density scaling exponent $a$, which in the original method could only be $2$ whereas in actual foam-like materials it can vary depending on the microscopic deformation mode.

These modifications allowed us to apply our model to mode-I fracture  of a highly porous brittle material under uniaxial compressive loading. The current method can accurately simulate nucleation of compressive 'anticracks' in areas of high stress concentration as well as their propagation, leading to cleavage of the material along the crack path. By using a short-range force approach for contact forces, we are able to simulate subsequent compaction of the material and demonstrate that the anticrack-type failure is followed by the gradual expansion of a compaction band in the direction perpendicular to the original failure plane. 

While the model allows control over the disorder in the system, it should be noted that using the current method a high value of disorder cannot be reached, as $p_{\phi,CV}$ quickly approaches zero for higher disorders. Therefore, further development of the model is required for studying the behaviour of highly disordered porous materials. Other developments, which are important for the modelling of porous materials such as snow where anticrack-type failure is of high practical importance, might include the new formation of cohesive bonds by sintering processes. 

\label{sec:conclusions}

\section*{Conflict of interest}
The authors declare that they have no conflict of interest.

\section*{Data availability}
Datasets generated and analyzed during this study are published and available on Zenodo, https://doi.org/10.5281/zenodo.15012216

\section*{Funding}
This research was funded by the Deutsche Forschungsgemeinschaft (DFG, German Research Foundation) - 377472739/GRK 2423/2-2023. The authors are very grateful for this support.

\bibliographystyle{spbasic}      
\bibliography{references_anticrack}   

\end{document}